# Variations of Pickup Ion Distributions and Their Relation to Interplanetary Conditions and Waves

Lukas Saul, Eberhard Möbius, and Charles W. Smith

*Department of Physics and Institute for the Study of Earth, Oceans, and Space;
University of New Hampshire, Durham, NH 03885*

**Abstract.** Pickup ion distributions vary substantially on a variety of time scales, although their sources may be relatively steady. This complicates their use as probes of the heliospheric and local interstellar particle populations. Interstellar He+ pickup ion observations from SOHO/CTOF and measurements of interplanetary conditions from SOHO and WIND enable a quantitative statistical analysis of these variations. Pickup ion distributions have been shown to correlate with IMF orientation, solar wind density, and IMF strength. Correlations of the pickup ions with IMF fluctuations are demonstrated, and it is shown that these are consistent with pitch angle scattering by waves. Further questions in pickup ion flux variations are discussed.

## INTRODUCTION

Interstellar neutral helium flows freely into the heliosphere, whereupon the dominant ionization process is solar EUV radiation [e.g. Rucinski, 1995]. This pickup ion (PUI) population is not only important for determining interstellar parmeters [e.g. Möbius et al., 1985, Möbius et al., 1995; Gloeckler et al., 1993] but also for analyzing plasma transport in the solar wind [e.g. Möbius et al., 1998; Chalov & Fahr 2000; Chalov & Fahr 2002; Saul et al., 2002].

Here we disccuss efforts to observe PUI transport in the inner heliosphere directly with in situ observations by SOHO/CELIAS/CTOF. This instrument enables high temporal resolution of singly charged helium in the thermal and suprathermal energy range [Hovestadt et al., 1995]. This PUI data is compared with bulk solar wind parameters from CELIAS/MTOF/PM and interplanetary magnetic field (IMF) data from WIND/MFI [Lepping et al., 1995].

## PICKUP ION TRANSPORT

The solar wind plasma is so dilute as to be effectively collisionless. After ionization or injection, PUIs are only acted upon by the electromagnetic fields embedded in the solar wind plasma. However, fluctuations in the IMF can act as scatterers and complicate matters. Some of the major resulting transport processes are listed in Table 1.

The most dominant force that diverts the PUIs from a free-streaming trajectory (and gives them their name) is the Lorentz force. A newly ionized interstellar atom is at motion with respect to the bulk solar wind plasma (and its frozen in magnetic field) and so feels an electric field $\vec{E} \sim \vec{v} \times \vec{B}$. Because the injection is a continuous process in a relatively homogenous medium, this produces a gyrotropic distribution. This ring distribution of PUIs injected into a solar wind plasma is shown in Figure 1 for the case of a perpendicular IMF.

When the IMF is not perpendicular, the resulting field felt by the injected pickup ions will be smaller, and the corresponding ring in velocity space will be smaller as well. The injected ring distribution is on the sunward hemisphere as seen in Figure 2. The resulting anisotropy is important as the measurements considered are mostly taken in the antisunward hemisphere.

## SCATTERING AND COOLING

This velocity distribution of newly ionized PUIs, known as a ring distribution, is non-Maxwellian and unstable to the generation of MHD waves [e.g. Lee & Ip, 1987; Zank & Cairns 2000]. The distribution is also subject to other wave-particle interactions, with any waves that are already present in the background solar wind. It is well established theoretically that such interactions will cause scattering, and diffusion of the distribution in pitch angle [e.g. Jokipii, 1972; Schlickeiser 1998, Chalov & Fahr 2002]. This pitch angle scattering is the second transport process on Table 1, and it acts to spread out the ring distribution on its spherical shell in velocity space.

| TABLE #1. Major Transport Processes acting on Helium Pickup Ions in the Solar Wind at 1AU | | |
|---|---|---|
| **Process Name** | **Description** | **Relevant Timescale** |
| Lorentz Force | Forms gyrotropic distribution. | ~10 sec. – 30 sec. |
| Pitch angle Diffusion | Forms isotropic distribution (due to magnetic fluctuations). | ~10 hrs. – 100 hrs. |
| Adiabatic Cooling | Expanding solar wind and decreasing IMF cools pickup ions. | ~100 hrs. |
| Energy Diffusion | Statistical acceleration by waves will scale with Alfven speed over solar wind speed. | ~1000hrs. |
| Coulomb scattering | SW density decreases as $r^{-2}$; mean free path becomes on order of size of the heliosphere. | ~50 days |

As the solar wind expands radially from the sun, the magnetic field decreases and the solar wind cools. The PUIs are also cooled in this process. The cooling acts to shrink the spherical shells in velocity space, so that cooling will slowly fill the full sphere [Vasilyunas & Siscoe, 1976]. Inner shells are more isotropic as they have spent more time susceptible to pitch angle scattering. In addition, energy diffusion may contribute to the spread of pickup ions across the shells in velocity space [e.g. Isenberg, 1987; Chalov & Fahr 2002].

Early models of PUI transport assumed that the time scale for pitch angle scattering is much quicker than for adiabatic cooling. However, scattering can be much slower, and the two processes were observed to act on a similar time scale in some cases [Gloeckler et al., 1995; Möbius et al., 1998]. These slower scattering rates have been discussed in terms of both lower turbulence levels and faster transit times (higher $V_{SW}$) [e.g. Chalov & Fahr, 2002].

In times of near radial IMF, the distribution starts off highly anisotropic (Figure 2). This anisotropy causes a reduced flux into the instrument in times of radial IMF [Möbius et al., 1998]. However, the anti-sunward hemisphere is filled more effectively during times of higher pitch angle scattering rates, e.g. higher wave power. In perpendicular fields, there is no initial sunward-antisunward anisotropy, and so scattering rates should not affect the observed PUI velocity distribution.

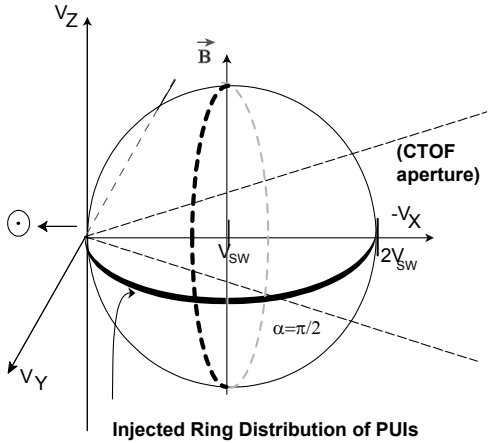
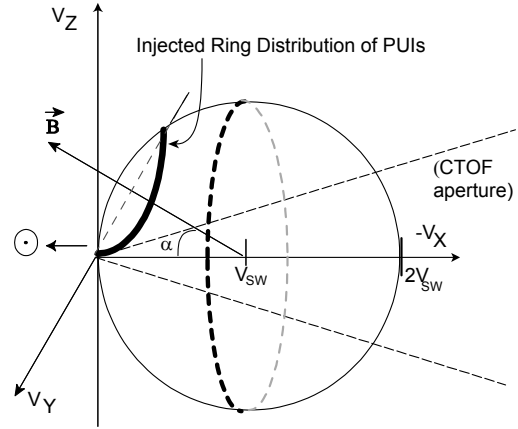

**FIGURE 1.** Schematic view of the velocity distribution of newly ionized pickup ions in a perpendicular magnetic field. Particles accessible to the CTOF instrument lie inside the dashed cone.

**FIGURE 2.** Schematic view of the velocity distribution of newly ionized pickup ions in an oblique magnetic field. Anisotropy in the X or radial component is now visible.

By comparing the pickup ion velocity distribution to the fluctuation spectrum of the IMF, we can test for wave-particle interaction effects in pickup ion transport. We applied Fourier analysis to 3 second magnetic field vectors from WIND/MFI (for further details see [Saul et al., 2004] and references therein). The resulting power spectra from 0.002 Hz to 0.16 Hz were fit to a power law. As a measure of resonant wave power, the value of this power law fit at 0.1 Hz (the helium gyrofrequency at ~ 14 nT) was used. The analysis was done for different orientations of IMF fluctuations. In particular, we consider here the power in fluctuations perpendicular (transverse) to the mean field, which we call $P_{TR}$. This index is a proxy for the power in Alfvenic fluctuations, which are thought to dominate pitch angle scattering.

## WAVES AND PITCH ANGLE ISOTROPIZATION

We report here a slightly different analysis of the wave-PUI interaction data than that of Saul et al. [2004], although the implications are similar. Observed fluxes near the cutoff velocity (from 1.9 to 2.1 times the solar wind speed in the spacecraft frame) are shown for times of radial IMF in Figure 3. Fifteen-minute flux averages were taken (y axis), and binned in the mean value of the transverse wave power (x axis) during those times. The error bars represent the statistical uncertainty of the mean for each transverse wave power bin. The visible trend of increasing flux near the cutoff with increasing resonant wave power is consistent with a decreased anisotropy due to pitch angle scattering by resonant waves.

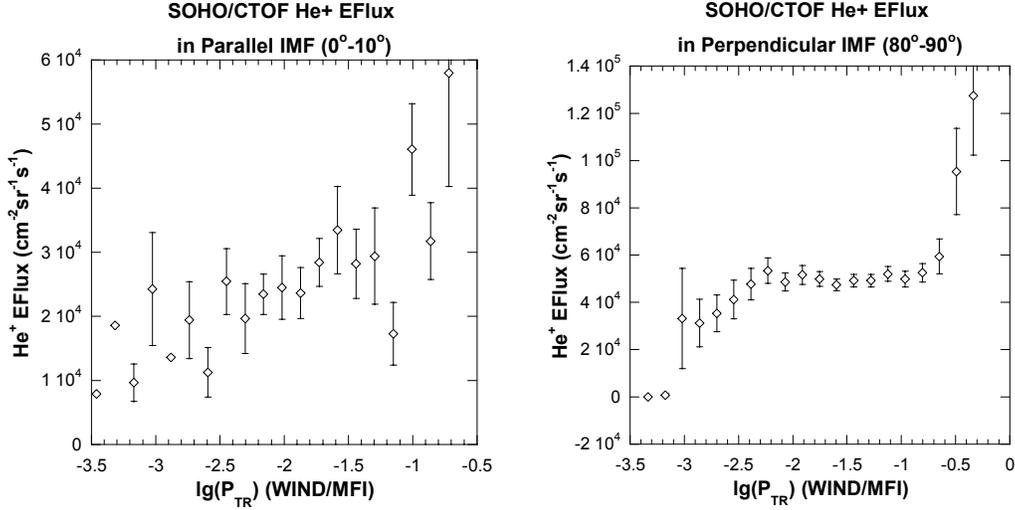

**FIGURES 3 & 4.** Average helium PUI energy fluxes during times of IMF near radial or parallel to $V_{SW}$ (left – Fig. 3) and IMF near perpendicular to $V_{SW}$ (right – Fig. 4) binned in different powers transverse resonant wave power. Only flux near the cutoff velocity ($1.9 < V/V_{SW} < 2.1$) is included.

For the case of perpendicular IMF (Figure 4), the amount of PUI flux is constant over most of the range of transverse wave power. This is as expected because PUI distributions are already effectively isotropic regardless of the scattering rate. However, in the highest range of wave power an increase in PUI flux is visible. In the context of wave-particle interactions this correlation is not well understood. It is also present in the similar analysis of Saul et al. [2004].

One possible explanation for the observed correlation of PUI flux and wave power during times of perpendicular IMF is an indirect or secondary correlation. It has been observed that PUI flux is correlated with proton density [e.g. Saul et al., 2002]. It is similarly seen (not shown here) that the SW proton density is also correlated with transverse wave power, so that more transverse waves are observed during times of high proton density as measured with MTOF/PM. These correlations can combine to yield a correlation of PUI flux with wave power, without an explicit causal relation. This could be a factor in producing the correlation of PUI energy flux near the cutoff and transverse wave power during times of perpendicular IMF (Figure 4). Such secondary correlations make interpretation of statistical correlations more difficult in general. This is especially true in the case of acceleration mechanisms, because during times of compressed or shocked SW, there is usually an associated increase in background magnetic field fluctuations or wave power.

# STOCHASTIC ACCELERATION

It has been suggested [e.g. Fisk, 1976] that energy diffusion of ion velocity distributions is enhanced by the presence of magnetic field fluctuations. A variety of theoretical work has been done on this problem of pickup ion transport by waves [e.g. Isenberg, 1987], but the experimental record in the solar wind remains thin, especially in comparison to the success of shock acceleration theories. Nonetheless, the observation of suprathermal tails in PUI populations even in quiet solar wind suggests that a 2$^{nd}$ order Fermi or stochastic acceleration mechanism is at work [Gloeckler, 2003].

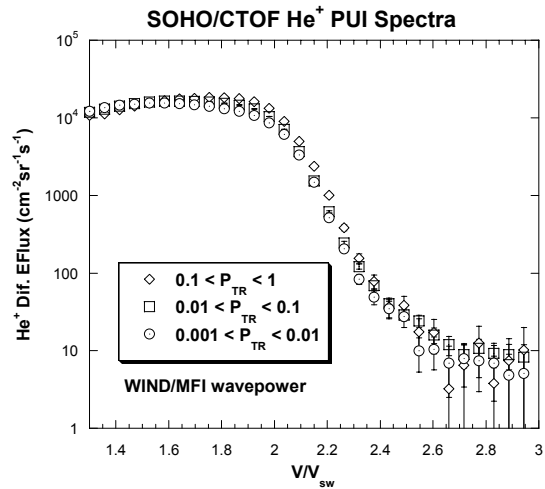

**FIGURE 5**. PUI spectra shown logarithmically to emphasize the suprathermal tail. The spectra are divided in three ranges of resonant transverse wave power. Time periods near published interplanetary shocks are not included.

To study this problem with the PUI test particle population, we examined logarithmic PUI flux spectra obtained by combining times of similar background wave power in a superposed epoch analysis. Times within 12 hours of a shock (as determined by the published WIND shock list) were excluded from the spectra. The results are shown in Figure 5, where three spectra are shown for different ranges of resonant transverse wave power.

No detectable difference in the suprathermal tails is present from one level of resonant wave power to another in this analysis. Other analysis of the parallel component of IMF fluctuations also indicated no observable correlation with PUI suprathermal tails (not shown here).

The implication is that resonant waves embedded in the solar wind (at 1AU in the ecliptic) do not appear to be the cause of PUI suprathermal tails. However, this analysis does not exclude other 2$^{nd}$ order Fermi acceleration processes such as non-resonant effects or scattering from lower frequency waves. Also a possibility is that resonant waves could do their work on a solar wind packet closer to the sun and yet not be observable at 1 AU. In this sense the analysis is complementary and not contradictory to [Schwadron et al., 1996], who found a statistical link with lower frequency IMF fluctuations.

# CONCLUSIONS

The presence of interstellar pickup ions in the solar wind at 1 AU is a boon to those who wish to study SW transport and wave-particle interactions. Here we have reported recent work which has found observational evidence of pitch angle isotropization by transverse waves. We have also put observational constraints on statistical acceleration mechanisms which are still not fully understood.

While the data are compelling we must emphasize that the dataset represents only solar minimum conditions at 1AU in the ecliptic. Further observations of PUI velocity distributions at other heliospheric distances and conditions as well as better modeling will greatly enhance our knowledge of PUI transport in the SW. Other correlations may emerge when sorting the data by transit time (solar wind speed), increasing the total time considered (statistics), or specifying the turbulent properties in more detail.

# ACKNOWLEDGMENTS

Helpful discussions with members of the UNH space science group including P. Isenberg, M.A. Lee, and Y. Litvinenko are gratefully acknowledged. This work is partially supported by NASA grants NGT5-50381, NAG5-10890, and NNG04GA24G, and by APL subcontract 872976.